%% file: aa17156.tex
\documentclass[structabstract]{aa}  

\usepackage{enumerate}
\usepackage{graphicx}
\usepackage{txfonts}
\usepackage{natbib}

\input{math_def.tex}

\begin{document}
  
  \title{MHD Dynamical Relaxation of Coronal Magnetic Fields}

  \subtitle{II. 2D Magnetic X-Points}

  \author{Jorge Fuentes-Fern\'andez, Clare E. Parnell and Alan W. Hood}

  \institute{School of Mathematics and Statistics, University of St Andrews, North Haugh, St Andrews, Fife, KY16 9SS, Scotland}

  \date{}

  \abstract
  {Magnetic neutral points are potential locations for energy conversion in the solar corona. 2D X-points have been widely studied in the past, but only a few of those studies have taken finite plasma beta effects into consideration, and none of them look at the dynamical evolution of the system. At the moment there exists no description of the formation of a non-force-free equilibrium around a two-dimensional X-point.}
  {Our aim is to provide a valid magnetohydrostatic equilibrium from the collapse of a 2D X-point in the presence of a finite plasma pressure, in which the current density is not simply concentrated in an infinitesimally thin, one-dimensional current sheet, as found in force-free solutions. In particular, we wish to determine if a finite pressure current sheet will still involve a singular current, and if so, what is the nature of the singularity.}
  {We use a full MHD code, with the resistivity set to zero, so that reconnection is not allowed, to run a series of experiments in which an X-point is perturbed and then is allowed to relax towards an equilibrium, via real, viscous damping forces. Changes to the magnitude of the perturbation and the initial plasma pressure are investigated systematically.}
  {The final state found in our experiments is a ``quasi-static'' equilibrium where the viscous relaxation has completely ended, but the peak current density at the null increases very slowly following an asymptotic regime towards an infinite time singularity. Using a high grid resolution allows us to resolve the current structures in this state both in width and length. In comparison with the well known pressureless studies, the system does not evolve towards a thin current sheet, but concentrates the current at the null and the separatrices. The growth rate of the singularity is found to be $t^D$, with $0<D<1$. This rate depends directly on the initial plasma pressure, and decreases as the pressure is increased. At the end of our study, we present an analytical description of the system in a quasi-static non-singular equilibrium at a given time, in which a finite thick current layer has formed at the null. The dynamical evolution of the system and the dependence of the final state on the initial plasma and magnetic quantities is discussed, as are the energetic consequences.}
  {}

  \maketitle


\section{Introduction}

Understanding the nature of hydromagnetic equilibria is a key issue for the study of solar magnetic fields such as those of the solar corona. Magnetic relaxation and field extrapolation are topics which clearly involve magnetic equilibria, and many other areas of study start from and/or aim to evolve to an equilibrium magnetic field, namely, magnetoacoustic wave propagation and dissipation, magnetic reconnection, or current sheet formation. However, the majority of these studies simply involve force-free equilibria, i.e. equilibria in which only magnetic forces are considered and forces due to plasma pressure, gravity, or velocity are all ignored.

In general, such assumptions are probably very reasonable in the majority of the low beta solar corona. However, in certain regions, such as near null points (where $|{\bf B}|$ is very small) and in the chromosphere, the effects of the plasma pressure become important. A study of the non-force-free relaxation of unbraided magnetic fields has started to be studied numerically and analytically in the first paper of this series \citep{Fuentes10}.

Magnetic energy release in the solar corona has for long been, and still remains, a key unresolved problem in solar physics. Magnetic null points play an important role in this release of magnetic energy, since they are likely to evolve by collapsing towards a tangential field discontinuity with a strong current, eventually leading to dissipation via magnetic reconnection. Such processes have been widely studied both analytically and numerically, with direct applications to solar environments such as in the CME breakout model \citep{Antiochos99}, which has been applied extensively in the last decade \citep[e.g.][]{Forbes06,Zucarello09}, and in other interplanetary scenarios such as the reconnection site in the Earth's magnetotail \citep[e.g.][]{Hesse01}. Also, they have been used in wave propagation experiments involving a zero beta plasma \citet{McLaughlin04} and a finite beta plasma \citep{McLaughlin06}, finding in both cases that the waves wrap around the null point, causing an exponential build up of current density at the location of the null. On the other hand, 2D reconnection has been studied for decades starting with \citet{Dungey53} and followed up by many \citep[e.g.][etc.]{Parker57, Sweet58, Petschek64, Biskamp86, Priest86, Craig94a, Shibata93}.

Furthermore, null points have been found to have a reasonably high population density in the solar corona, by \citet{Longcope09}. In two dimensions, in particular, the topology of the current sheet formed after an X-type null point collapse has been well studied analytically for potential and force-free fields by several authors \citep[e.g.][]{Dungey53,Green65,Somov76,Craig94, Bungey95}.

The aim of this paper is to provide a valid magnetohydrostatic equilibrium from the collapse of a two-dimensional X-point, which are simpler scenarios than 3D null points, although the latter will be studied in a follow-up paper. Under ideal, non-resistive conditions, the energy bound up in the global magnetic field has to manifest itself as localized accumulations of current density.

It is well known that under the cold plasma approximation (e.g. zero plasma beta), an initially perturbed X-point field relaxes to a potential equilibrium with a Y-type infinitesimally thin current sheet where the current is zero everywhere except within the magnetic tangential discontinuity, where it develops a singularity of the form $j_z=\delta(A_z-A_{z0})$. These potential configurations are described by \citet{Green65} and \citet{Somov76}. Later, \citet{Bungey95} expanded these solutions for potential and force-free fields giving a general expression for these force-free current sheets. Latter studies have found the formation of localised infinite current layers in the Earth's magnetotail \citep{Birn03}, relevant for the initiation of the subsequent energy release phase.

\citet{Friedel96} and \citet{Grauer98} studied two-dimensional current singularities in incompressible flows using an Adaptive Mesh Refinement (AMR) code which permitted them to obtain very highly localised resolutions at the locations of maximum current. They found an exponential growth for the current at the null point.

An analytical result by \citep{Klapper97} has rigorously shown that, in 2D ideal incompressible plasmas, a singularity of the current density will take an infinite amount of time to develop, unless driven by a pressure singularity occurring outside a neighborhood of the null point. In this paper we aim to show numerically that this is also true for compressible plasmas with a finite plasma pressure.

First evidence of current sheets extending along the separatrices in sheared magnetic field structures were studied by \citet{Zwingmann85} for force-free equilibria, where they found mathematical singularities in the current sheet which they interpreted as terms that ``would become large in a real physical situation''. Later, \citet{Vekstein93} made a mathematical analysis of the magnetic field around cusp-points, after the shearing of a magnetic field with an X-type null point, and suggested a form of a negative power law for the resulting singular current density as a function of $A_z$.

\citet{Rastatter94} considered for the first time the effects of pressure perturbations in numerical experiments on the ideal relaxation of two-dimensional magnetic X-points, and studied the development of current layers with singular current densities in which, in the relaxed state, the initial X-point was replaced by either a T-point or a cusp-point geometry. For the relaxation, they used a frictional code, which damped the kinetic energy out of the system by adding a fictitious relaxing term to the momentum equation of the form $-\kappa{\bf v}$, but without any associated heating term in the energy equation. Their X-point relaxed to a singular equilibrium with a plasma pressure jump across the separatrices, and with current layers extending along the separatrices. They argued that the finite width of their current sheet was due to the finite difference method in their numerical approach rather than being real, but found, nevertheless, the integrated current density over the sheet width (referred as to surface current) to be constant along each whole separatrix.

Later, \citet{Craig05} reconsidered the problem of the relaxation of two-dimensional magnetic X-points and the formation of current singularities in non-force-free equilibria, and evaluated the strength of the current singularity at the end of their relaxation. Again, they made use of a frictional code with a fictitious damping term, $-\kappa{\bf v}$, added to the momentum equation, but with no heating term in the energy equation, assuming the polytropic model $p\sim\rho^{\gamma}$, which imposes a condition of adiabaticity to the process. In analogy with the results of \citet{Rastatter94}, they found a distribution of current density extended along the magnetic separatrices, which they claimed to be almost uniform. They repeated their study for a number of different grid resolutions, and presented a logarithmic increase of the peak current with the number of grid points, at the same time as the area of the current layer above a given value for the current showed a logarithmic decrease. Hence, the current layer itself became narrower with higher resolution.

Then, they evaluated the scalings of the peak current for different values of the background plasma pressure of the system, finding a weakening of the growth of the peak current density as the plasma pressure was enhanced. That is, a singularity is harder to achieve the higher the value of the plasma pressure, although the presence of a non-zero plasma beta would not prevent a singularity forming. The results of \citet{Craig05} were fully reproduced by \citet{Pontin05}, as part of their study of current singularities at 2D and 3D nulls, using a Lagrangian code, instead of the Eulerian code used in \citet{Craig05}.

Here, we study the dynamical evolution of magnetic X-points under the frozen-in assumption, which is driven by viscous forces which involve a heating term that ensures energy conservation. The forces of gravity are neglected, as the pressure scale-heights in the solar corona (of the order of 100Mm) are far larger than our length-scales, here understood (given our line-tied boundaries) as the height of the possible magnetic null points above the base of the corona, which might be of the order of 1Mm \citep[see][]{Longcope09}. We show how a type of singularity wants to be formed in the non-force-free case, in agreement with the numerical studies of \citet{Rastatter94} and \citet{Craig05}. Our numerical results show how the initial X-point collapses to a cusp-like geometry in which the current density accumulates around the neutral point and along the four separatrices. Again, the results agree, in this aspect, with the previous numerical works of non-force-free X-point collapse. However, we attempt to go a step further by looking at the time evolution of the field, running a series of very high resolution experiments, which allow us to look closer at the current accumulations. Also, following the lines of \citet{Vekstein93}, we propose a mathematical form for the final equilibrium state, involving a finite current layer ending in cusp-points at both extremes. 

The paper is structured as follows: In Sec. \ref{sec:sec2}, we present the initial setup and the details of the numerical experiments. In Sec. \ref{sec:sec3} we state the equations that define our final 2D magnetohydrostatic equilibrium. The results from the numerical experiments are analysed closely in Sec. \ref{sec:sec4}, whilst in Sec. \ref{sec:sec5} we comment on the nature of an analytical description for the equilibrium field. Finally, we conclude with a general overview of the problem in Sec. \ref{sec:sec6}.


\section{Numerical scheme and initial setup}\label{sec:sec2}

For the numerical experiments studied in this paper, we have used Lare2D, a staggered Lagrangian-remap code with user controlled viscosity, that solves the full MHD equations, with the resistivity set to zero \citep[see][]{Arber01}. In order to create the initial magnetic field, a current-free hyperbolic X-point, $A_z=(x^2-y^2)/2$, is perturbed by squashing it in the vertical $y$-direction by a given amount $(1-h)$ times the height of the original system, without introducing any plasma flow, such that the flux function of the initial state is given by
\begin{equation}
A_z(x,y,0)=\frac{1}{2}\left(x^2-\frac{y^2}{h^2}\right)\;. \label{ini_Az}
\end{equation}
The squashing creates a uniform non-zero current density whose $z$-component is
\begin{equation}
j_z(x,y,0)=\frac{1}{h^2}-1\;. \label{ini_jz}
\end{equation}
The squashing is characterised by the height of the box $h$, normalised to the original height, so that the dimensions of the experiment are $L\times hL$. The initial plasma pressure $p_0$, is set to a constant everywhere, such that the initial system is not in equilibrium.

The numerical domain is a 2D box with a uniform grid of $1024\!\times\!2048$ points. The higher resolution in the $y$-direction is chosen to permit any current layer that may form to be as thin as possible, but still resolvable as far as possible across its width.

Magnetic field lines are line-tied at the four boundaries and all components of the velocity are set to zero on the boundaries. The other quantities have their derivatives perpendicular to each of the boundaries set to zero. Hence, quantities that are conserved over the whole domain are total energy and total mass. Since the process is ideal (there is no diffusion to within the numerical limits), the field is frozen to the plasma, and mass in a single flux tube (or along a field line) must be conserved. The integrated current density is also conserved throughout the dynamical evolution. This can be easily shown, taking the symmetry properties of the system into account,
\begin{eqnarray}
\frac{{\rm d}}{{\rm d}t}\int_S\!{\bf j}\cdot{\bf ds}&=&\frac{1}{\mu}\frac{{\rm d}}{{\rm d}t}\int_S\!\boldnabla\times{\bf B}\cdot{\bf ds} \nonumber\\
&=&\frac{1}{\mu}\frac{{\rm d}}{{\rm d}t}\oint_C\!{\bf B}\cdot{\bf dl} \nonumber\\ 
&=&\frac{1}{\mu}\frac{{\rm d}}{{\rm d}t}\oint_C\!B_t{\rm d}l\,=\,0\;.
\end{eqnarray}
Here, $S$ represents the surface of our numerical domain, with ${\bf ds}$ being a vector perpendicular to it, $C$ is the the curve that encloses $S$, and ${\bf dl}$ is a vector tangent to $C$ at each point. The integral of the tangential magnetic field, $B_t$, along the curve $C$, which is the boundary of our box, equals zero for each pair of boundaries (left-right and top-bottom), because of symmetry.

The numerical code uses the normalised MHD equations, where the normalised magnetic field, density and lengths,
\begin{eqnarray*}
x=L\hat{x}\;,\;\;\;y=L\hat{y}\;,\;\;\;{\bf B}=B_n\hat{\bf B}\;,\;\;\;\rho=\rho_n\hat{\rho}\;,
\end{eqnarray*}
imply that the normalising constants for pressure, internal energy and current density are
\begin{eqnarray*}
p_n=\frac{B_n^2}{\mu}\;,\;\;\;\epsilon_n=\frac{B_n^2}{\mu\rho_n}\;\;\;{\rm and}\;\;\;j_n=\frac{B_n}{\mu L}\;.
\end{eqnarray*}
The subscripts $n$ indicate the normalising constants, and the {\it hat} quantities are the dimensionless variables with which the code works. The expression for the plasma beta can be obtained from this normalization as
\begin{eqnarray*}
\beta=\frac{2\hat{p}}{\hat{B}^2}\;.
\end{eqnarray*}
In this paper, we will work with normalised quantities, but the hat is removed from the equations for simplicity.

The (normalised) equations governing our MHD dynamical processes are
\begin{eqnarray}
\frac{\partial \rho}{\partial t}+\boldnabla\cdot(\rho{\bf v}) &=& 0\;,\label{n_mass}\\
\rho\frac{\partial{\bf v}}{\partial t}+\rho({\bf v}\cdot\boldnabla){\bf v} &=& -\boldnabla p + (\boldnabla\times{\bf B})\times{\bf B} + {\bf F}_{\nu}\;,\label{n_motion}\\
\frac{\partial p}{\partial t}+{\bf v}\cdot\boldnabla p &=& -\gamma p \boldnabla\cdot{\bf v}+H_{\nu}\;,\label{n_energy}\\
\frac{\partial{\bf B}}{\partial t} &=& \boldnabla\times({\bf v}\times{\bf B})\;,\label{n_induction}
\end{eqnarray}
where ${\bf F}_{\nu}$ and $H_{\nu}$ are the terms for viscous force and heating, and internal energy is given by the ideal gas law, $p=\rho\epsilon(\gamma-1)$, with $\gamma=5/3$.

We have run a number of experiments with various heights ranging from $h=0.9$ to $h=0.6$, with the subsequent initial current densities ranging from $j_0=0.23$ to $j_0=1.78$, and various initial plasma pressures ranging from $p_0=1.0$ to $p_0=0.125$. Following a systematic study, we find that the function $\mathcal{F}(A_z)$ in equations (\ref{p_Az}) and (\ref{j_Az}) differs from one experiment to another, and is therefore determined by the initial plasma pressure and current density of the system.


\section{Magnetohydrostatic equilibrium in 2D}\label{sec:sec3}

In order to understand, and be able to give a description of, the final equilibrium state of our numerical experiments, we will now describe the basic equations of a two-dimensional MHS equilibrium.

In the absence of gravity, flows or frictional forces, the momentum equation reduces to the following magnetohydrostatic (MHS) equation,
\begin{equation}
{\bf j}\times{\bf B} - \boldnabla p = 0\;, \label{equi}
\end{equation} 
where ${\bf j}$ and ${\bf B}$ are the electric current density vector and magnetic field vector, respectively, and $p$ is the plasma pressure. For our 2D problem, all the quantities depend on the horizontal and vertical coordinates, namely $x$ and $y$. Making use of Amp\`ere's law, $\boldnabla\times{\bf B}=\mu_0{\bf j}$, the 2D equilibrium can be characterised by the {\it Grad-Shafranov equation},
\begin{equation}
\frac{dp}{dA_z} = -\frac{1}{\mu_0}\nabla^2A_z \;, \label{gradshaf}
\end{equation}
which indicates that, in equilibrium, {\it the plasma pressure is a function of the flux function, $A_z$}, where ${\bf B}=\boldnabla\times(0,0,A_z(x,y))$, so such a state is uniquely determined by a function of the form
\begin{equation}
p=\mathcal{F}(A_z)\;, \label{p_Az}
\end{equation}
where $\mathcal{F}$ is an unknown function that is dependent on the initial conditions and evolution. Since, in 2D, the magnetic field lines are defined by contours of $A_z$, the Grad-Shafranov equation tells us that in a MHS equilibrium, {\it the plasma pressure is constant along field lines}.

In addition, it can easily be shown that $\nabla^2A_z=-j_z$, where $j_z$ is the $z$-component of the electric current density. Hence, $j_z$ should be also characterised by a unique function of $A_z$,
\begin{equation}
j_z=\mathcal{F}^{\prime}(A_z)=\frac{d\mathcal{F}}{dA_z}\;, \label{j_Az}
\end{equation}
when the system is in equilibrium, so that {\it current density is also constant along field lines}.

\begin{figure*}[t]
  \begin{minipage}[b]{1.0\linewidth}
    
    \begin{minipage}[b]{0.49\linewidth}
      \centering
      \includegraphics[scale=0.32]{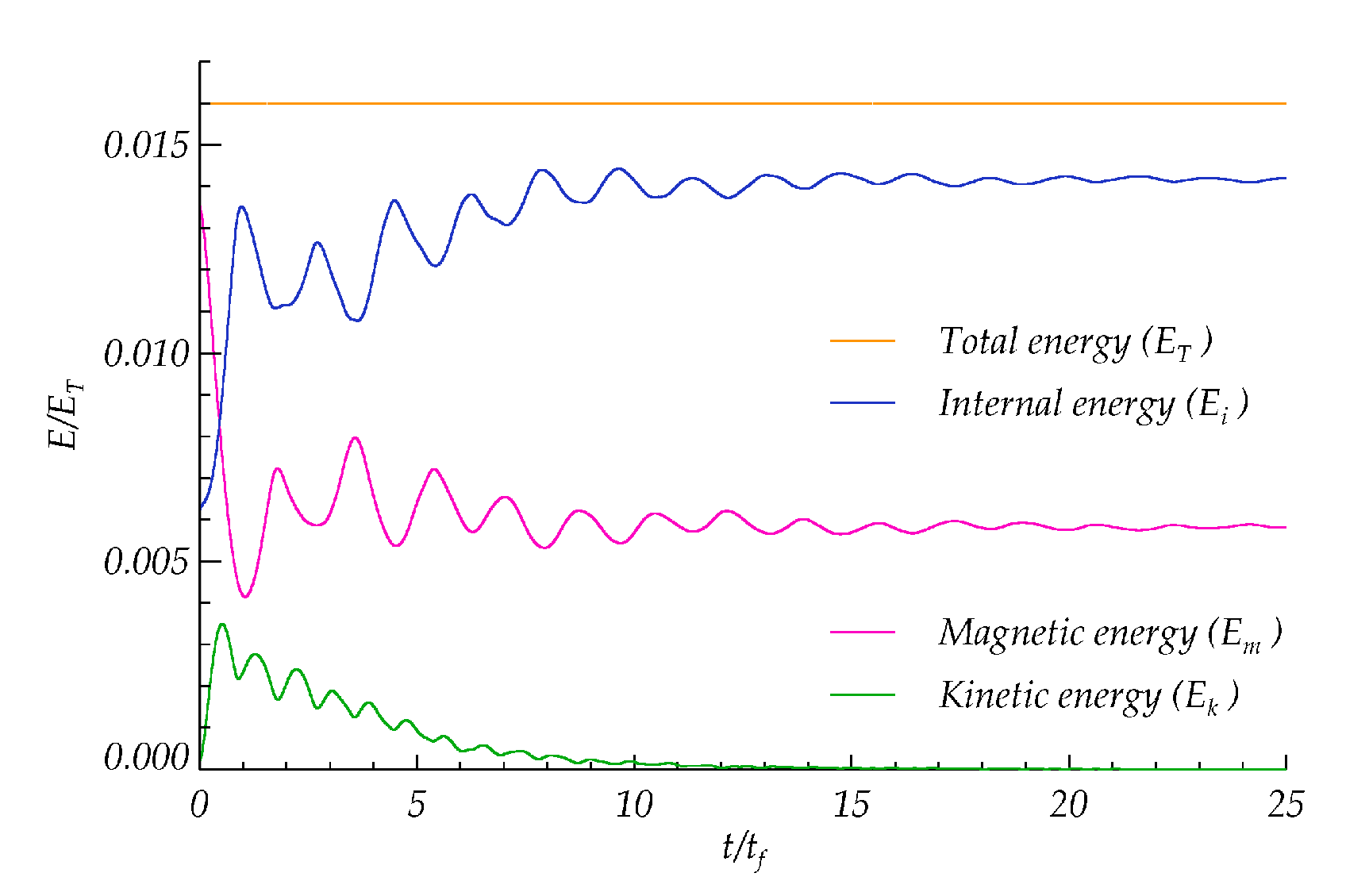}
      \caption{Time evolution of the energies of the system, integrated over the whole two-dimensional box, for an experiment with $j_0=1.04$ ($h=0.7$) and $p_0=0.250$. The magnetic, internal and total energies have been shifted on the y-axis by subtracting the values 0.238, 0.742 and 0.984 respectively, but their amplitudes are not to scale.}
      \label{fig:energy}
    \end{minipage}
    \hspace{0.02\linewidth}
    \begin{minipage}[b]{0.49\linewidth}
      \centering
      \includegraphics[scale=0.32]{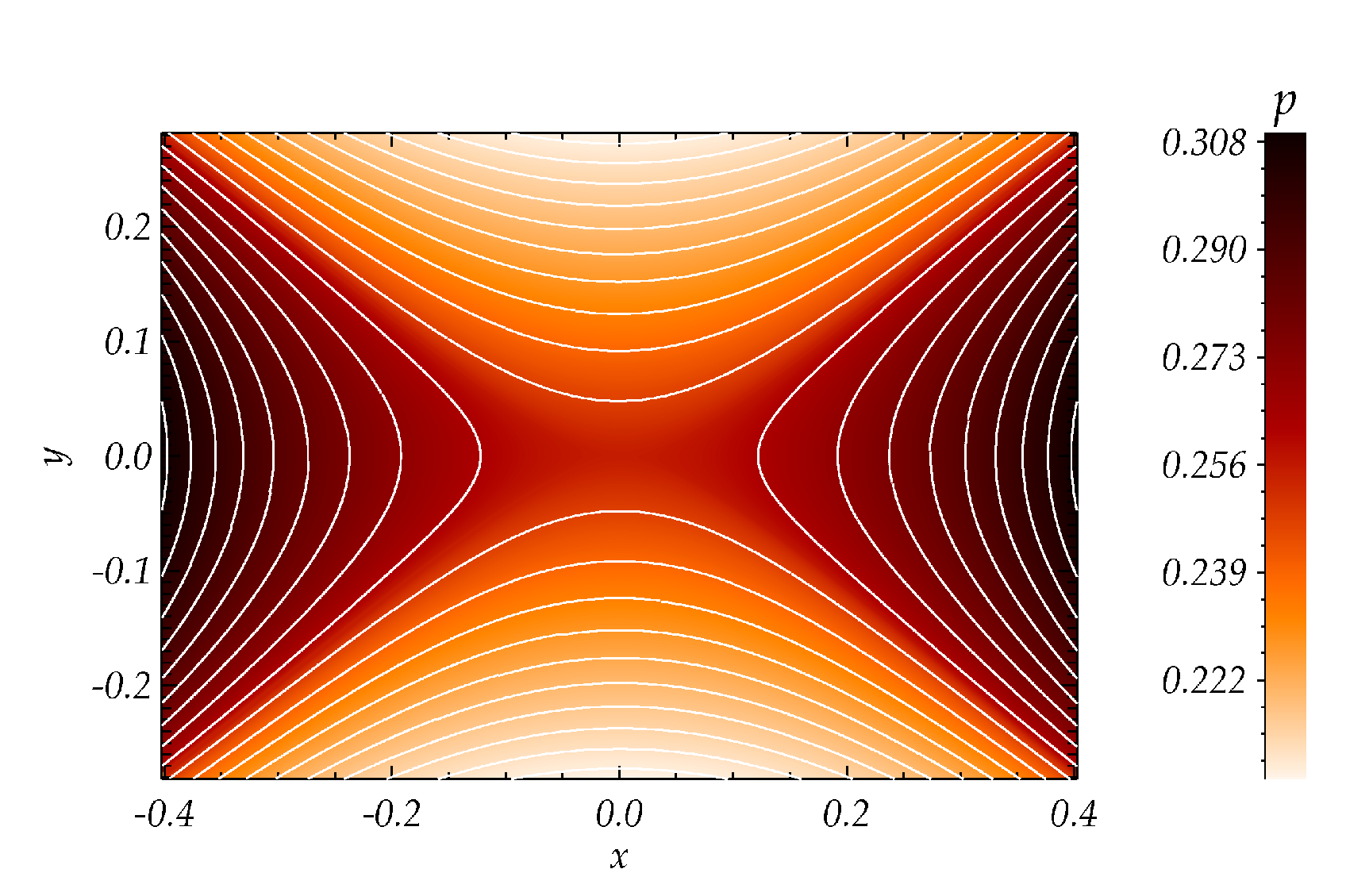}
      \caption{Two-dimensional contour plot of plasma pressure for the final equilibrium state for the sample experiment with $h=0.7$ and $p_0=0.250$. White solid lines are the magnetic field lines as contours of the flux function $A_z$.}
      \label{fig:2Dpressure}
    \end{minipage}
    
    \vspace{0.3cm}
   
  \end{minipage}
  \begin{minipage}[b]{1.0\linewidth}
    
    \begin{minipage}[b]{0.5\linewidth}
      \centering
      \includegraphics[scale=0.32]{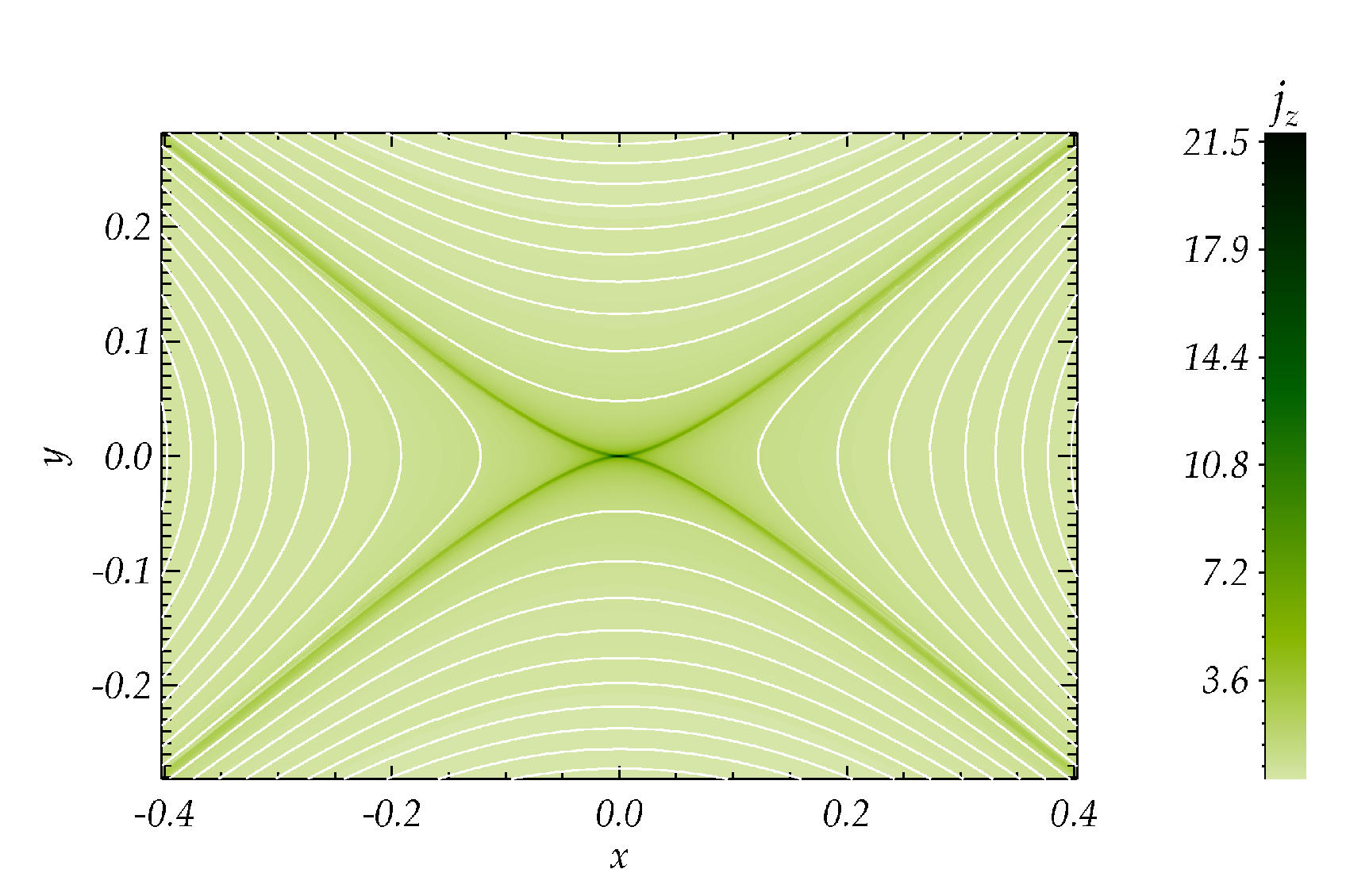}
    \end{minipage}
    \begin{minipage}[b]{0.5\linewidth}
      \centering
      \includegraphics[scale=0.32]{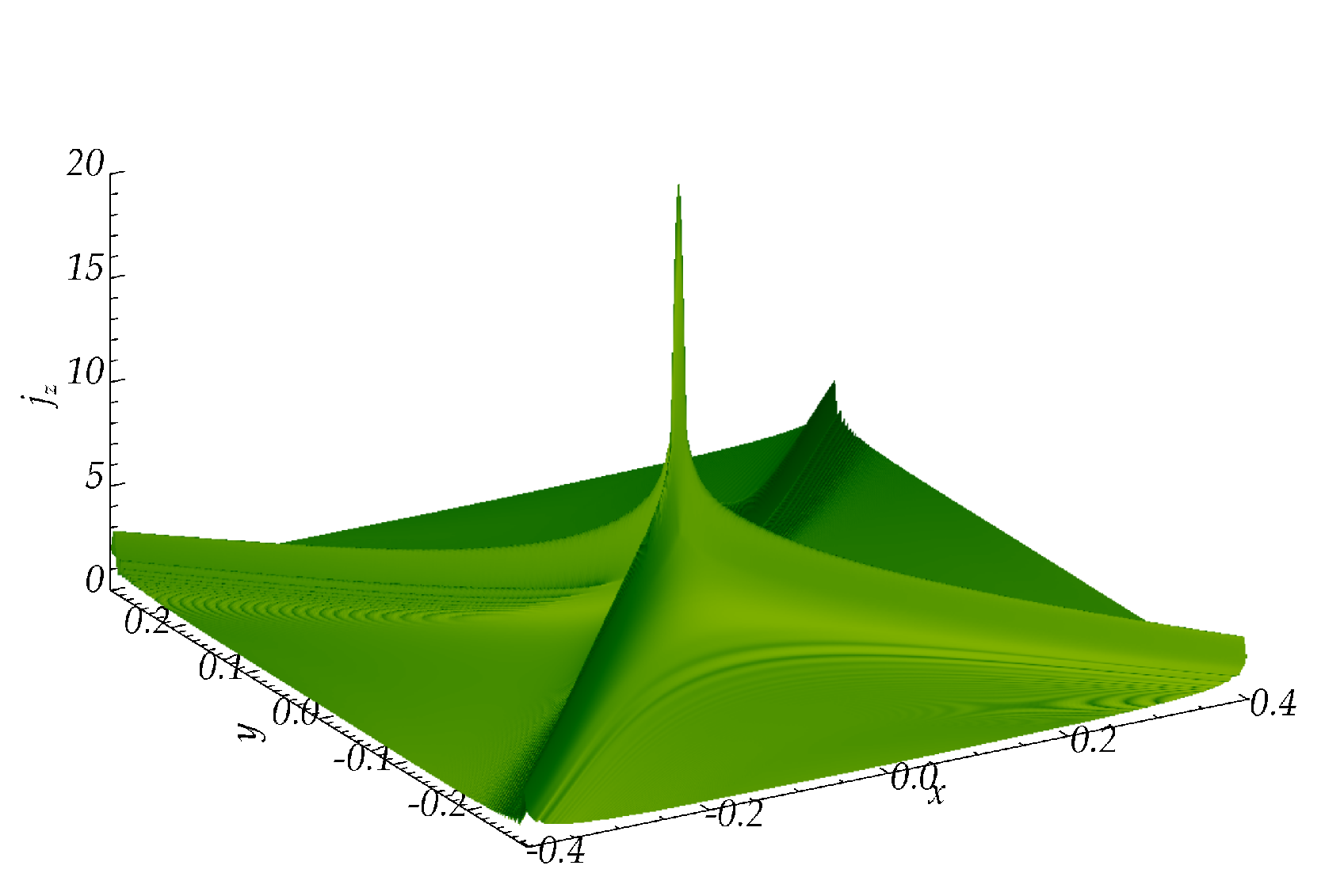}
    \end{minipage}
    
    \caption{Two-dimensional contour plot (right) and surface (left) of electric current density for the final equilibrium state for the sample experiment with $h=0.7$ and $p_0=0.250$. White solid lines are the magnetic field lines as contours of the flux function $A_z$.}
    \label{fig:2Dcurrent}
  \end{minipage}
  \vspace{0.3cm}
\end{figure*}


\section{Numerical Results}\label{sec:sec4}


\subsection{Energetics}

Initially, the only force acting on the system is the magnetic Lorentz force. This acts on the plasma by rising its kinetic energy, which is zero initially. The velocities are then damped out through the viscous forces, and the physical MHD evolution allows the transfer of the kinetic energy into internal energy (heat), via the viscous heating term. This is a key process in the time evolution of the system, as it ensures the conservation of total energy, and hence, it will play a role in the final state.

Fig. \ref{fig:energy} shows the time evolution of the four energies of the system: kinetic, magnetic, internal and total, integrated over the whole two-dimensional box, as a function of the computation time normalised to the fast magnetoacoustic time, $t_f$, defined as the time for a fast magnetoacoustic wave to reach the center of the box from the bottom or top boundary.

The overall period of oscillation of the magnetic energy is about two fast magnetoacoustic time units, while the kinetic energy, which is proportional to the velocity squared, has a period of oscillation of one fast magnetoacoustic time unit. The frequency ratio of these two oscillations is consistent with the Fourier analysis of the system, which involves a factor of two for the frequency associated to the internal energy.

At the final equilibrium, where kinetic energy goes to zero, there has been an exchange from magnetic to internal energy. Magnetic energy is converted to kinetic energy through the momentum equation, and then this kinetic energy is converted into internal energy (heating) through the viscous heating term. Total energy is conserved throughout the whole process.

\begin{figure*}[t]
  \begin{minipage}[b]{1.0\linewidth}
    
    \begin{minipage}[b]{0.49\linewidth}
      \centering
      \includegraphics[scale=0.45]{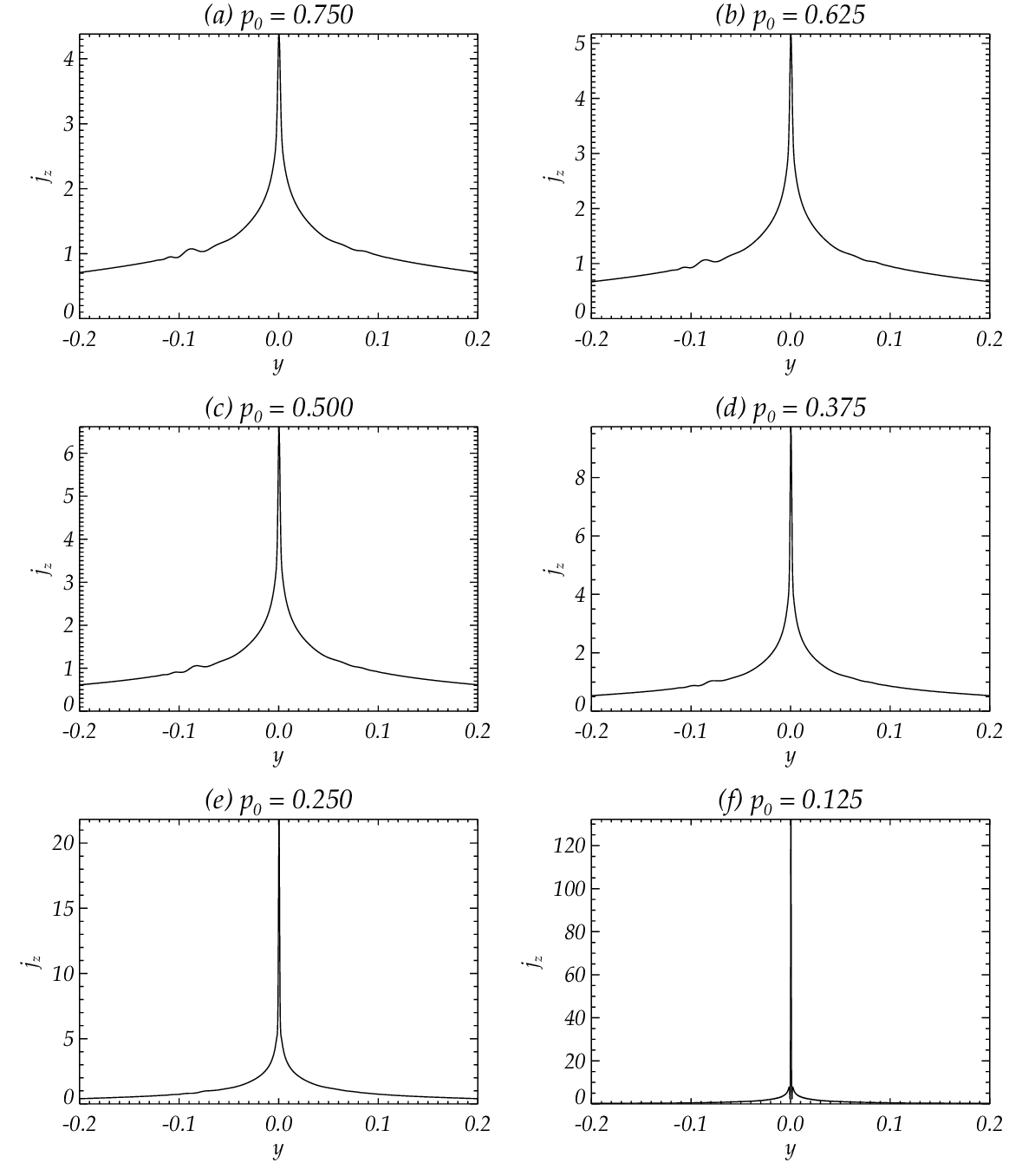}
      \caption{Plots of electric current density across the width of the central current layer, for six different experiments, with $h=0.7$, but different initial plasma pressures.}
      \label{fig:width}
    \end{minipage}
    \hspace{0.02\linewidth}
    \begin{minipage}[b]{0.49\linewidth}
      \centering
      \includegraphics[scale=0.45]{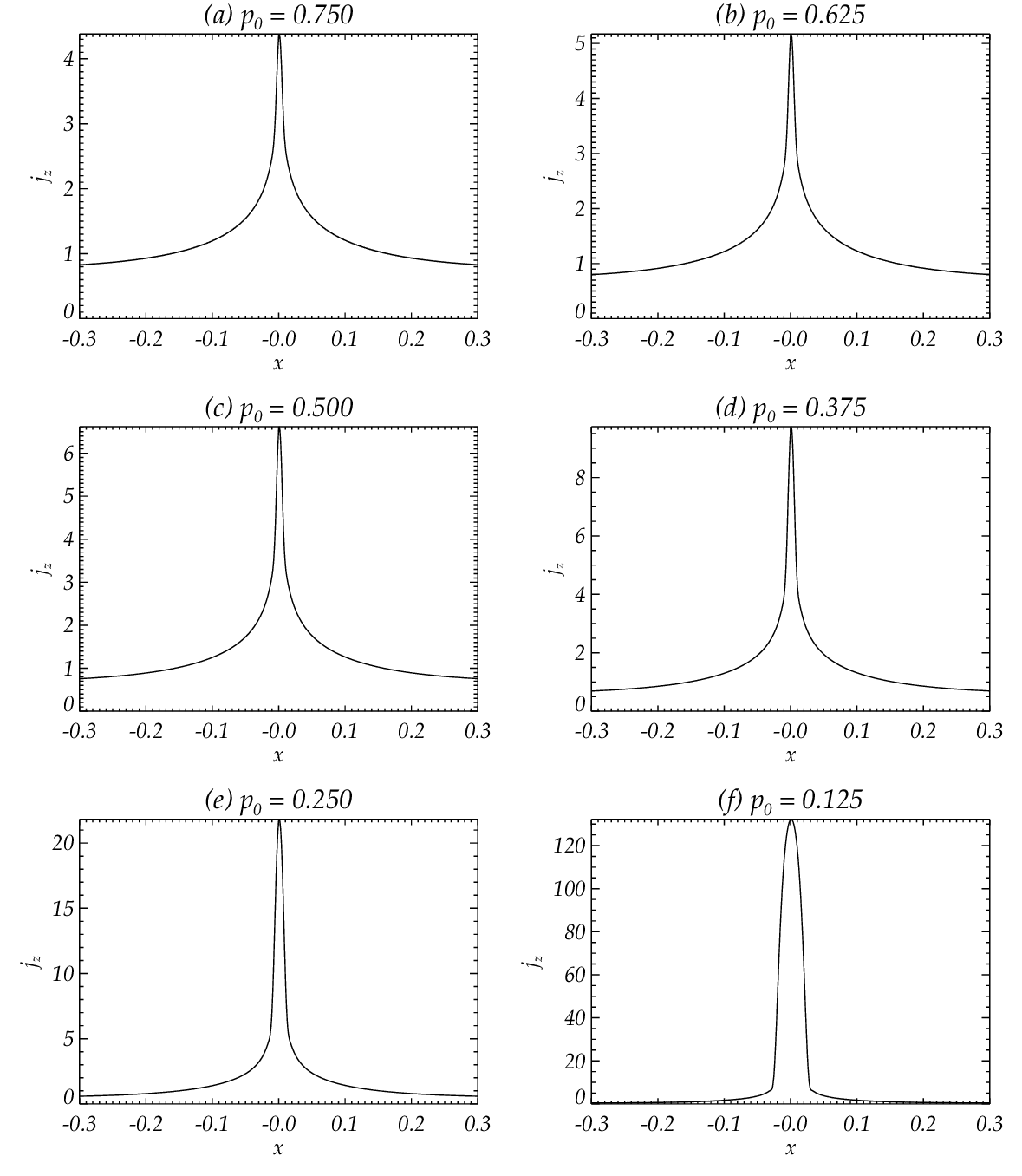}
      \caption{Plots of electric current density along the length of the main current sheet, for six different experiments, with $h=0.7$, but with different initial plasma pressures.}
      \label{fig:sheet}
    \end{minipage}
    
    \vspace{0.3cm}
    
  \end{minipage}
  \begin{minipage}[b]{1.0\linewidth}

    \centering
    \includegraphics[scale=0.65]{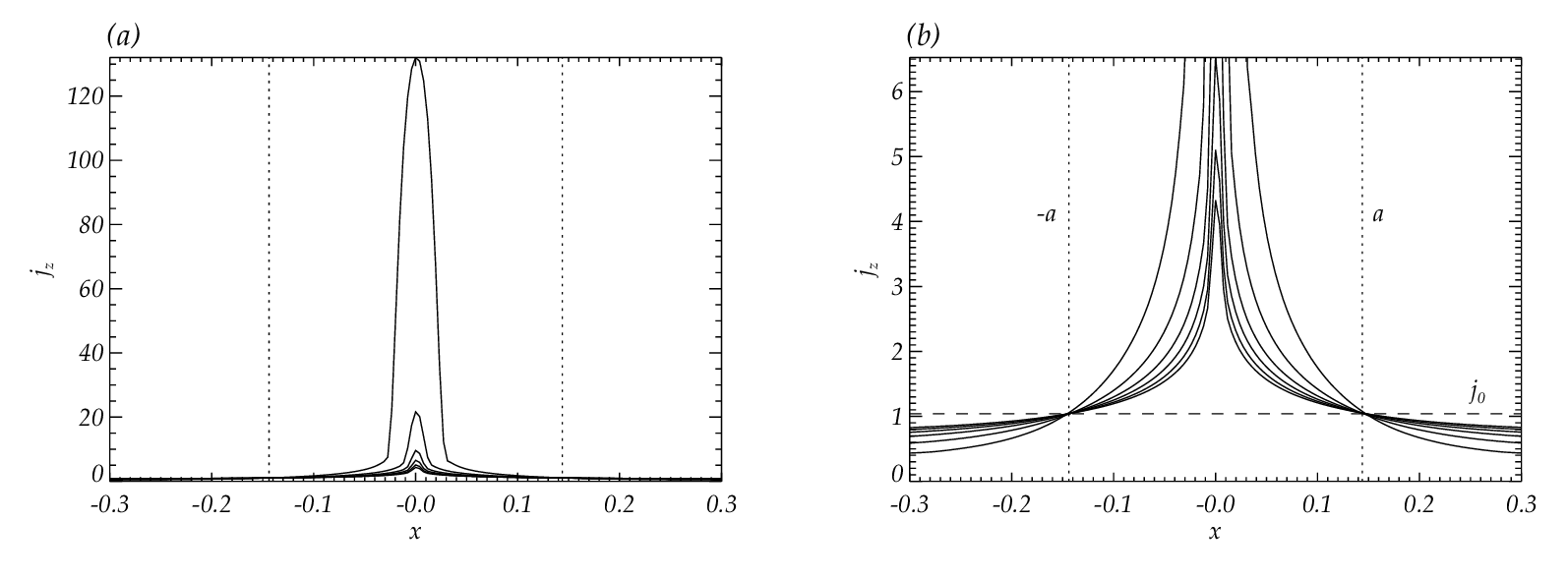}
    \caption{The six plots in Fig. \ref{fig:sheet} are overplotted for comparison. The dimensions of the Green's potential solution are given by the dotted lines. Figure (b) is a zoom of (a) over a smaller range of current densities. In (b), the initial current density is overplotted (dashed).}
    \label{fig:sheet2}

  \end{minipage}
  \vspace{0.3cm}
\end{figure*}


\subsection{Final state}

In Figures \ref{fig:2Dpressure} and \ref{fig:2Dcurrent}, we show two-dimensional contour plots of the plasma pressure and electric current density, respectively, in the final state (defined here at $t=100$, or equivalently, about 290 fast-magnetoacoustic-wave transits), for an experiment with $j_0=1.04$ ($h=0.7$) and $p_0=0.250$, together wit a three dimensional surface of the current density. Departing from an initial state containing an X-point with uniform pressure and current density, we get an equilibrium where the X-point has produced a thick current layer from which arms of enhanced current extend along the curved separatrices (Fig. \ref{fig:2Dcurrent}). The separatrices form cusp shapes at the two ends of the current layer. The plasma pressure is enhanced within the cusps (to the left and right of the current layer), and decreased in the regions outside the cusp, above and below the current layer (Fig. \ref{fig:2Dpressure}). Plasma pressure appears to be constant along field lines. The electric current density is clearly not constant along the separatrices, but the surface current -current integrated over two given field lines at each side of the separatrix- is checked to be constant along them.

We now look closely at the dimensions of the current layer for six experiments with $j_0=1.04$ ($h=0.7$) but different initial plasma pressure. In Fig. \ref{fig:width}, we show vertical cuts of the current density across the central current layer. The width of the central layer decreases for smaller plasma pressures, but remains finite. In Fig. \ref{fig:sheet}, we show horizontal cuts of the current density along the central current layer, for the same cases as in Fig. \ref{fig:width}. As the pressure is decreased, the length of the central current layer extends further, and the current density becomes more concentrated, developing a higher peak. This behaviour can be also observed if the initial plasma pressure is held fixed, and the height of the box is systematically decreased (i.e. the squashing is increased). Decreasing the initial plasma pressure has a similar effect as increasing the initial current density, as the action of both is to make the Lorentz force dominate over the pressure force.

When the initial plasma pressure is small (e.g. in Fig. \ref{fig:sheet}f), the current layer has a length that is many times longer that its width. We consider whether the current layer is approaching the form found in Green's current sheet solution. This is checked by comparing these plots with the length of Green's sheet (Fig. \ref{fig:sheet2}), given, by comparison with \citet{Bungey95}, as $a=0.25\sqrt{j_0/\pi}$, which for the sample experiment in Fig. \ref{fig:sheet2}, with $j_0=1.04$ ($h=0.7$), gives the value $a\approx0.144$.

In Fig. \ref{fig:sheet2}, the six horizontal cuts of Fig. \ref{fig:sheet} are overplotted, and the dimensions of Green's potential solution are marked. All the curves cross at the same points on $x$ and $y$, namely $(\pm a,j_0)$, corresponding to the initial value of the current density in $y$, and the two ends of Green's current sheet in $x$. The main conclusion that may be extracted from these plots is that the field is in all cases very far from the potential solution, although the fact that all curves cross at the ends of Green's potential sheet seems to imply that Green's solution might be achieved (as far as we can get with the resolution) in the limit $p_0\to0$.


\subsection{Current singularity}

\begin{figure}[ht]
  \centering
  \includegraphics[scale=0.58]{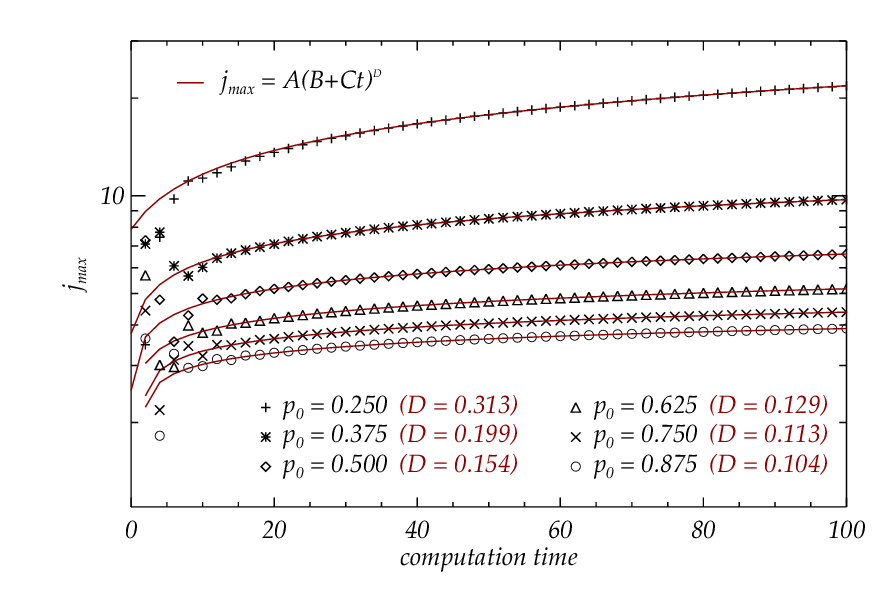}
  \caption{Magnitude of the electric current density at the location of the null, as a function of time, for six different experiments, with $h=0.7$, but with different initial plasma pressures.}
  \label{fig:peak1}
\vspace{0.3cm}
  \includegraphics[scale=0.58]{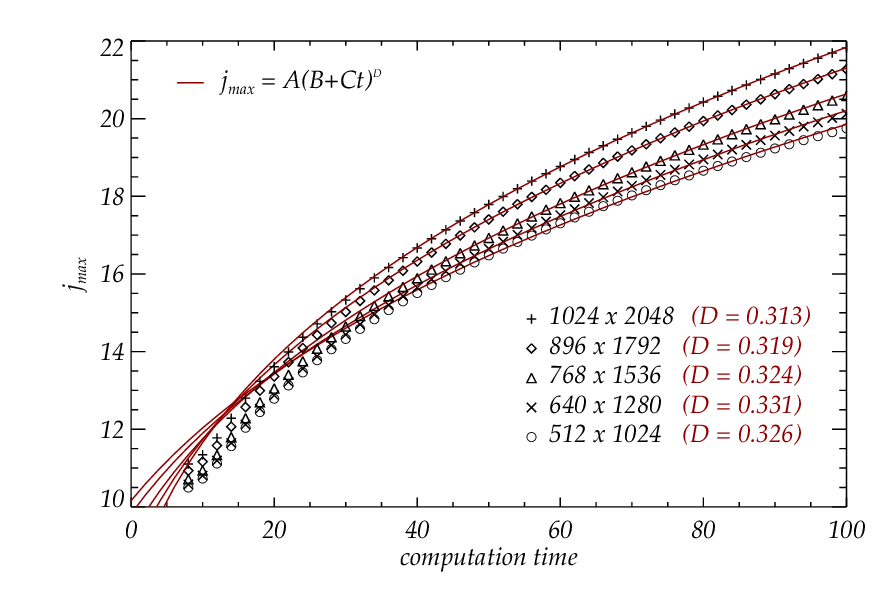}
  \caption{Magnitude of the electric current density at the location of the null, as a function of time, for the same experiment with $h=0.7$ and $p_0=0.250$, but with different grid resolutions.}
  \label{fig:peak2}
\vspace{0.3cm}
\end{figure}

The system has now reached a quasi-static equilibrium, and it is in an asymptotic regime where the current at the location of the null and along the separatrices slowly grows as time elapses. To evaluate the formation of a singularity we must look closer at the evolution of the current at the location of the null, namely, the ``peak current''. In Fig. \ref{fig:peak1} we show the time evolution of the peak current, $j_{max}$, for six different experiments with different initial plasma pressures. The first part of the evolution (from $t=0$ to $t\sim25$) consists of the propagation of magnetoacoustic waves, bouncing back and forward between the boundaries and in the magnetic separatrices, which are quickly damped out by the viscous forces. After this relaxation has finished, the field is in equilibrium everywhere outside the separatrices, and the peak current at the null follows a very slow time evolution of the form
\begin{equation}
j_{max}=A(B+Ct)^D\;,
\end{equation}
with $D>0$. Both the maximum current, and the exponent $D$ decrease as the plasma pressure is increased. That is, if a singularity is being formed, a large plasma pressure hinders the formation of such \citep[as discussed by][]{Craig05}. In any case, the formation of a singularity is a slow process that occurs in an infinite time, and, hence, it is impossible to reach, no matter how good the numerical resolution is. The key point here is that, once that the asymptotic regime has been achieved, any further grid improvement will only permit a longer run, but will not change the overall results and conclusions. These results are in agreement with \citet{Klapper97}, who rigorously showed that for 2D ideal incompressible plasmas, a singularity cannot form in a finite time. In Fig \ref{fig:peak2}, we show the evolution of the peak current for the same experiment but with different grid resolutions. It is shown that for both the value of the peak current, $j_{max}$, and the growth rate, $D$, the effect of varying the grid resolution at this level is small.

It is worth noting that MHD resistive instabilities such as the tearing mode \citep{Furth73} may occur, halting the relaxation and dissipating the current layer. In the idealised cases presented in the literature, the onset of this instability is independent of the resistivity and will occur, theoretically, for all values of $\eta$, when the length-to-width ratio of the current layer has gone below $2\pi$. It is found that our final quasi-equilibrium state is close to the condition given by this very simplified analytical model. However, we have carefully checked that for all the results shown, no significant numerical reconnection has taken place at the location of the null, since the changes of the flux function at the origin remain below a relative value of $10^{-6}$.

The study of such singularities is a complex topic, as they are, by definition, impossible to reach. A good way of approaching this problem is by using mesh-refinement codes \citep[e.g.][]{Friedel96,Grauer98} in which the grid resolution can be finely adjusted at certain locations. Nevertheless, these methods are not required for our purposes in this paper, as our high fixed resolution permits us to fully resolve the current accumulations both at the null point and along (and across) the four separatrices (see Sec. \ref{sec:layer}). Moreover, it has been shown that the evolution of the singularity does not change its behaviour when different grid resolutions are used.

\onecolumn
\begin{figure*}
  \centering
  \includegraphics[scale=0.65]{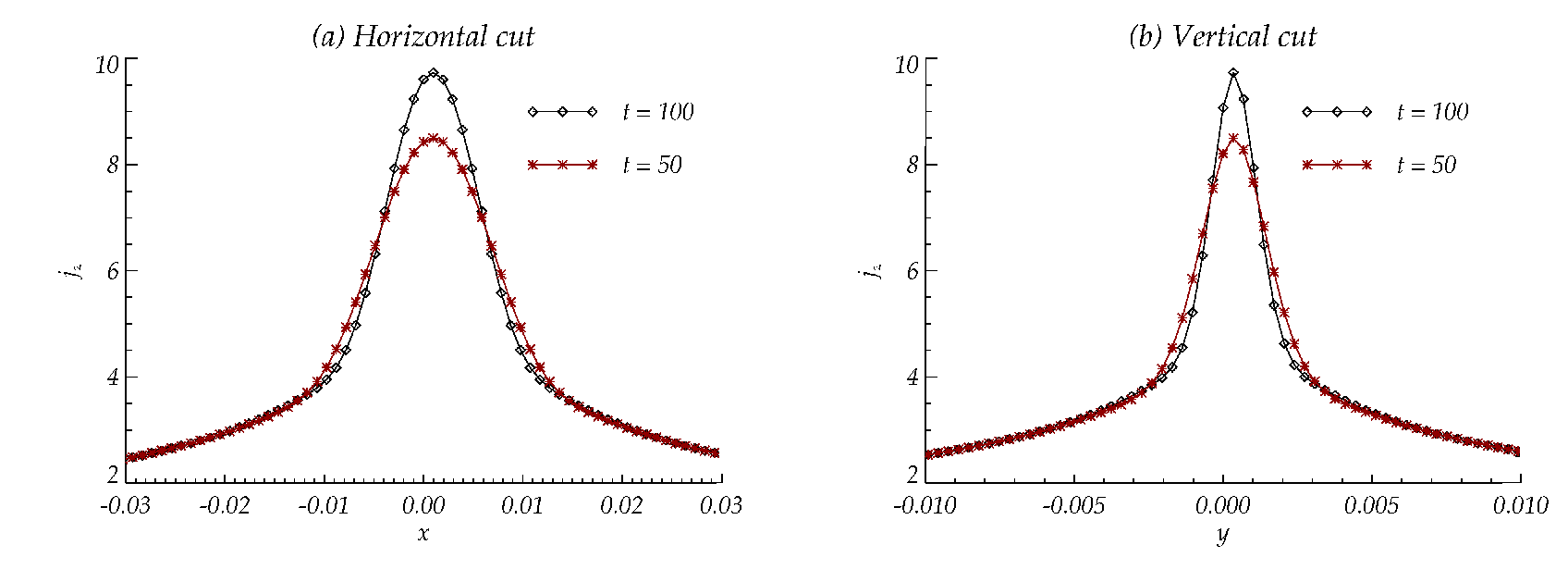}
  \caption{Electric current density across the width and along the length of the central current layer, for the same experiment with $h=0.7$ and $p_0=0.375$, at two different times, $t=100$ (black diamonds) and $t=50$ (dark red stars). These examples have the resolution of  $1024\times2048$.}
  \label{fig:dimensions}
  \vspace{1.0cm}
  \centering
  \includegraphics[scale=0.65]{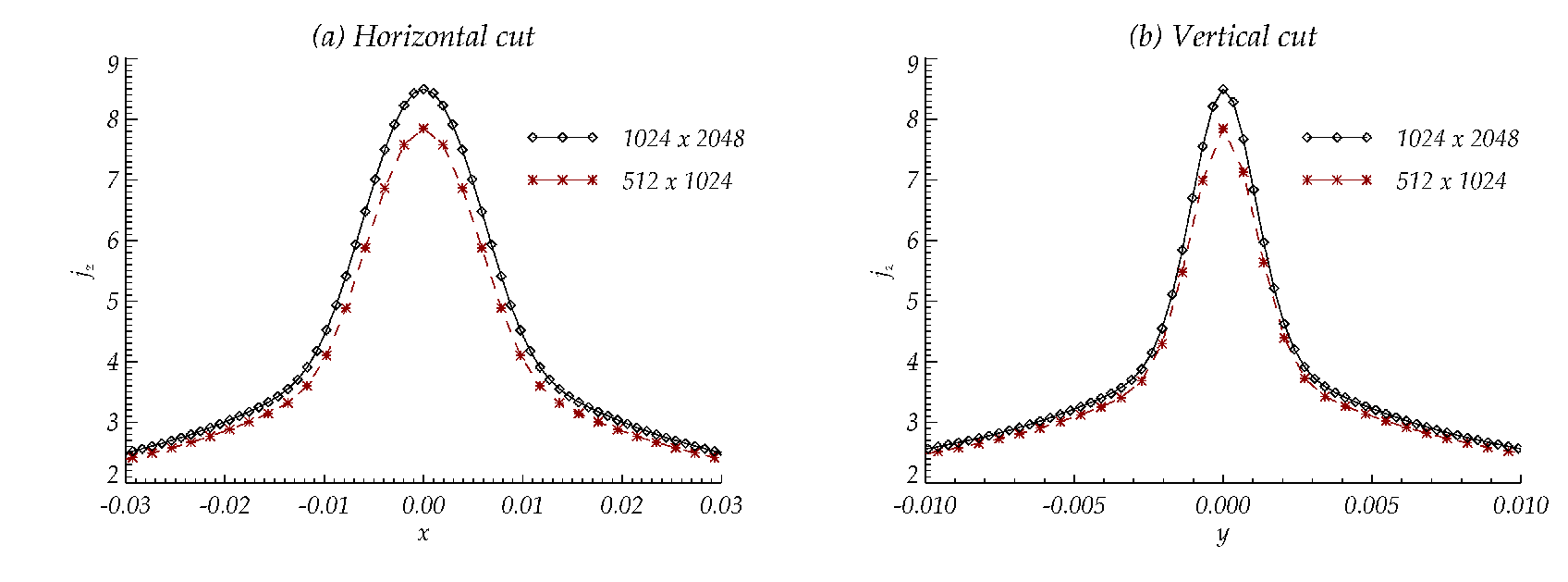}
  \caption{Electric current density across the width and along the length of the central current layer, for the same experiment with $h=0.7$ and $p_0=0.375$, at the time $t=50$. Results from the high resolution run, $1024\times2048$ (black diamonds) are compared with the results of a half-resolution run, $512\times1024$ (dark red stars).}
  \label{fig:resolution}
  \vspace{1.0cm}
  \centering
  \includegraphics[scale=0.65]{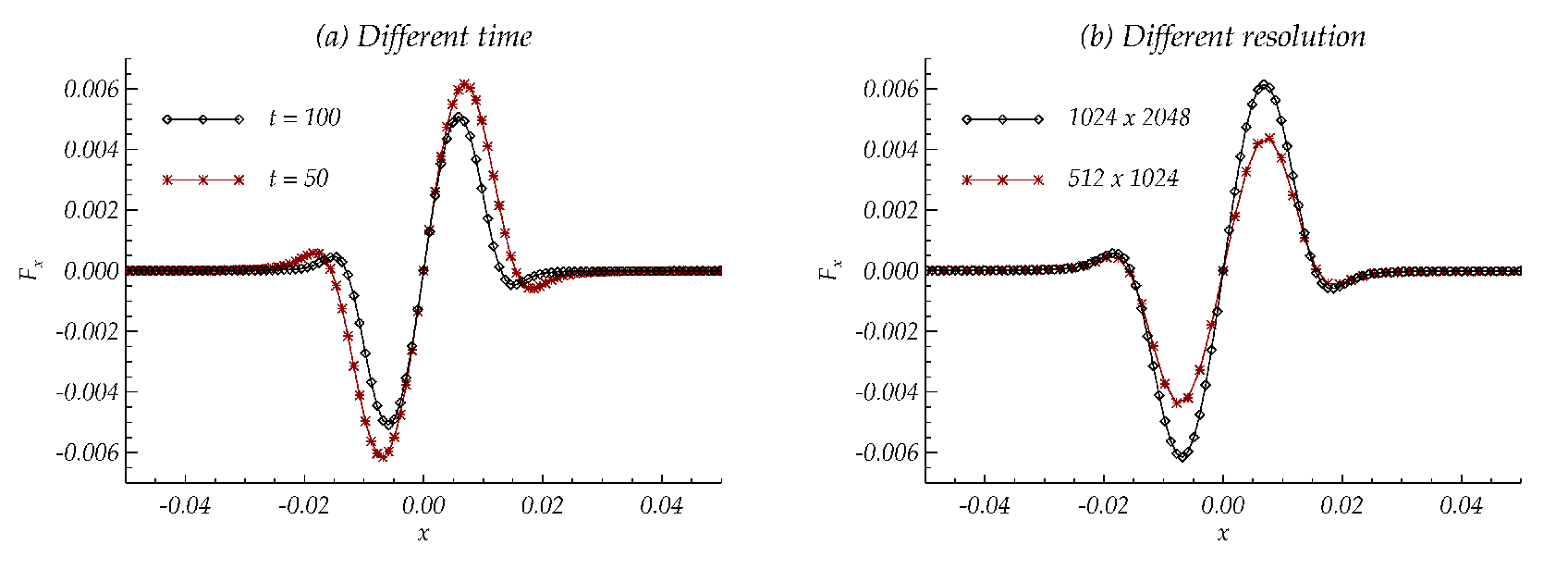}
  \caption{Residual forces along the central current layer, for the same experiment with $h=0.7$ and $p_0=0.250$, at (a) two different times, $t=100$ (black diamonds) and $t=50$ (dark red stars), and (b) with two different resolutions, $1024\times2048$ (black diamonds) and $512\times1024$ (dark red stars).}
  \label{fig:forces}
\end{figure*}
\twocolumn


\subsection{Current layer} \label{sec:layer}

The discussed on-going formation of an infinite time singularity suggests that small residual forces are acting at the location of the null, and hence, that the current layer is not in a perfect static equilibrium. In Fig. \ref{fig:dimensions} we look at the dimensions of the central current layer at different times. Our high resolution experiments allow us to resolve the central layer both in length and width. It is shown here that both the length and the width of the layer get smaller as time elapses.

Also, we check the dimensions of the current layer at equal times but with different grid resolutions (Fig. \ref{fig:resolution}), finding that while the peak current varies slightly with resolution, the dimensions of the current layer (at half the maximum value) {\it do not} vary.

\begin{figure}[t]
\centering
\includegraphics[scale=0.60]{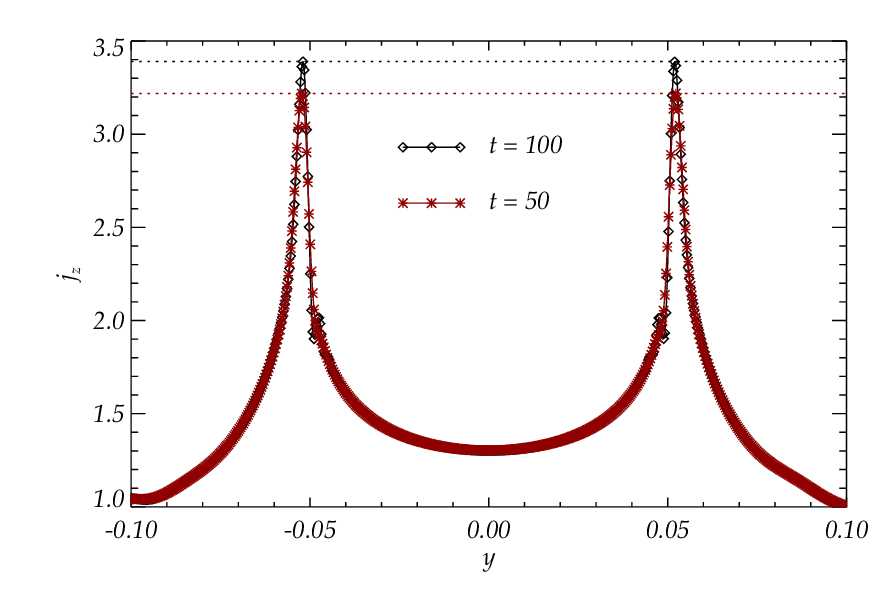}
\caption{Vertical cut of the current density at $x=0.1$ for the experiment with $h=0.7$ and $p_0=0.375$, at two different times, $t=100$ (black diamonds) and $t=50$ (dark red stars), with the resolution of  $1024\times2048$. The horizontal dotted lines at the top indicate the maximum currents in this cut.}
\label{fig:separ}
\vspace{0.3cm}
\end{figure}

Finally, we look at the residual forces at the location of the null. These are checked to be zero everywhere in the domain except for the region very closed to the null that we show here. Fig. \ref{fig:forces} shows the $x$ component of the total force, i.e. ${\bf j}\times{\bf B}-\boldnabla p$, along the length of the central current layer, at different times and with different resolutions. It is shown that forces become smaller with time, but also they concentrate in a smaller region, meaning that the current layer gets shorter with time.

As for the resolution, the regions in which the forces act (which we identify with the length of the current layer) stays the same, but the amplitude of the forces increases with resolution. This is a sign that these forces are not due to numerical reconnection, but are due to a singularity being formed.

In conclusion, the system seems to be forming a singularity, but the process is slow enough so that we can describe it as a quasi-static equilibrium at a given time, in which the current layer at the location of the null has a finite length, a finite width and a finite amplitude. As time gets very large, the current density layer gets smaller and thinner, but the peak value increases. It can be shown (Fig. \ref{fig:separ}) that the current along separatrices also increases with time, as the length of the central layer decreases. This may suggest that this current layer is feeding current into the separatrices.

Our aim now is to look for an analytical description of the field about the null in this quasi-static equilibrium at a given time, that would precede reconnection in a non-ideal environment.


\section{Analytical description of the field}\label{sec:sec5}


\begin{figure}[t]
  \centering
  \includegraphics[scale=0.60]{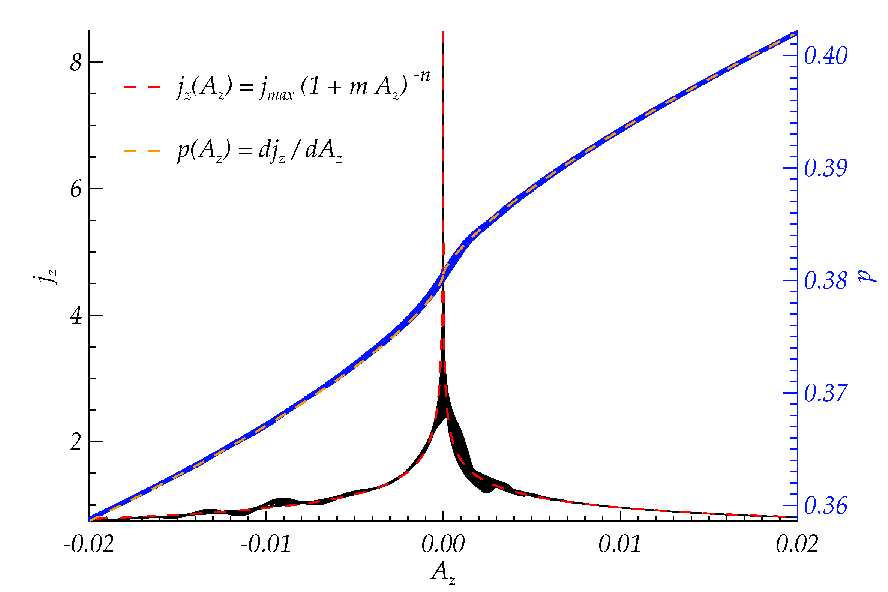}
  \caption{Test of Grad-Shafranov condition for the experiment with $h=0.7$ and $p_0=0.375$. Current density (black, y-axis on the left) and plasma pressure (blue, $y$-axis on the right) are plotted against the flux function $A_z$, for every single point in the numerical domain. Positive values of $A_z$ refer to inside of the cusp, while negative $A_z$ are outside the cusp. The functional form $j_z(A_z)=j_{max}(1+mA_z)^{-n}$ is overplotted here.}
  \label{fig:gradshaf}
  \vspace{0.3cm}
  \includegraphics[scale=0.60]{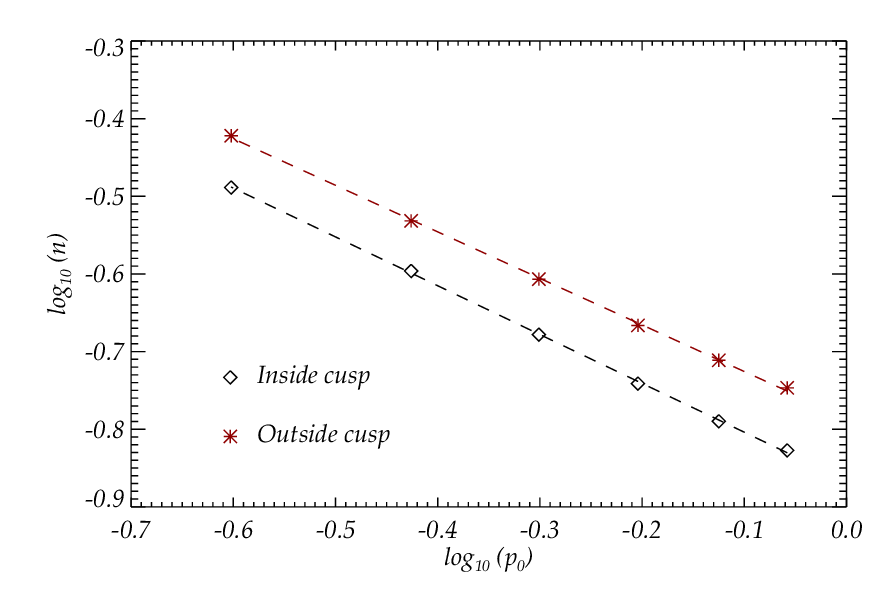}
  \caption{Logarithmic plot of the exponent $n$ for both regions inside and outside the cusp, as a function of the initial pressure, for a set of six experiments with the same squashing $h=0.7$, after the same time has elapsed. The exponents follow a power law with a very similar decay rate for the two regions. The slopes are $-0.63$ and $-0.6$ respectively for inside and outside the cusp.}
  \label{fig:parameters}
  \vspace{0.3cm}
\end{figure}

\subsection{Grad-Shafranov equation}

A direct check on the validity of our equilibrium may be done by testing the behaviour of the pressure $p$ with respect to the flux function $A_z$, and also the consequences of the Grad-Shafranov condition, (\ref{gradshaf}), which states that the current density $j_z$ must be also a unique function of $A_z$. We follow the approach given by \citet{Vekstein93} in an attempt to give a mathematical description of the field about the null in our quasi-static equilibrium. They suggested the form $j_z(A_z) = m A_z^{-n}$, with $n>0$, which is singular at the location of the null, where $A_z=0$. The problem is that the state that we want to describe is not exactly singular at the null (even if it is slowly converging towards a singularity). Therefore, we try a function of the form
\begin{equation}
j_z(A_z) = j_{max}(1+m A_z)^{-n}\;, \label{fit}
\end{equation}
where $j_{max}$ is the value of the peak current at the null at a given time, and $n>0$. This function is such that when $mA_z\gg 1$, the field behaves as if it were singular at the origin, with the form
\begin{equation}
j_z(A_z)=\frac{j_{max}}{m^n}A_z^{-n} \;, \label{sing}
\end{equation}
but when $mA_z\ll 1$, then $j_z=j_{max}$. The value of $m$ needs to be very large to counteract the small values of $A_z$ about the null. Note, that the separatrices do not satisfy the form given in Eq. \ref{fit}, as $A_z=0$ but $j_z\ne j_{max}$ along them.

In Fig. \ref{fig:gradshaf}, we represent {\it every single point} of the two-dimensional domain, for the plasma pressure and the current density against the flux function. It appears clear from this graph that the pressure is a unique function of $A_z$, and so is most of the current density distribution. The biggest dispersion occurs when we approach $A_z=0$, which is the value on the separatrices, and at the X-point, where a singularity may be being formed, which, from the Grad-Shafranov equation, implies an infinite derivative of the pressure with respect to $A_z$.

On Fig. \ref{fig:gradshaf}, we have overplotted the best fit to Eq. (\ref{fit}) to the current density, using $j_{max}=8.5$, and overplotted its derivative with respect to $A_z$ to compare to the plasma pressure. The best fit values, independently determined for the regions inside and outside the cusp, are:
\begin{eqnarray}
m_i&=&55\times10^4 \; \nonumber\\
n_i&=&0.253 \, \nonumber\\
m_o&=&17\times10^4 \; \nonumber\\
n_o&=&0.294 \;, \nonumber
\end{eqnarray}
where the subscripts $i$ and $o$ denote inside and outside the cusp, respectively. It is clear that the form of the equilibrium is different for the regions inside and outside the cusp. These fits can be done for all the experiments with different initial plasma pressures or initial current densities. In Fig. \ref{fig:parameters}, the exponents $n$ are plotted for different initial plasma pressures for both inside and outside the cusps. These exponents follow in both cases a negative power law with a very similar slope, suggesting, once more, that an increase in plasma pressure acts against the formation of a singular current density.


\subsection{Poloidal flux function}

\citet{Vainshtein90} proposed a description of the field about special magnetic points in two-dimensions, such as cusp points at the ends of thin current sheets, seeking a solution at small $r$ of the form
\begin{equation}
A_z(r,\theta)=r^{\alpha_1}g_1(\theta)+r^{\alpha_2}g_2(\theta)+...+r^{\alpha_n}g_n(\theta) \;, \nonumber
\end{equation}
with $1\le\alpha_1<\alpha_2<...<\alpha_n$, thus avoiding $A_z\to\infty$ when $r\to 0$. Here, $r$ is the radial coordinate whose origin is at the location of the null, and $\theta$ is the angular coordinate which is zero at the main axis of the current layer (i.e. $x$-axis).

\begin{figure}[t]
\centering
\includegraphics[scale=0.52]{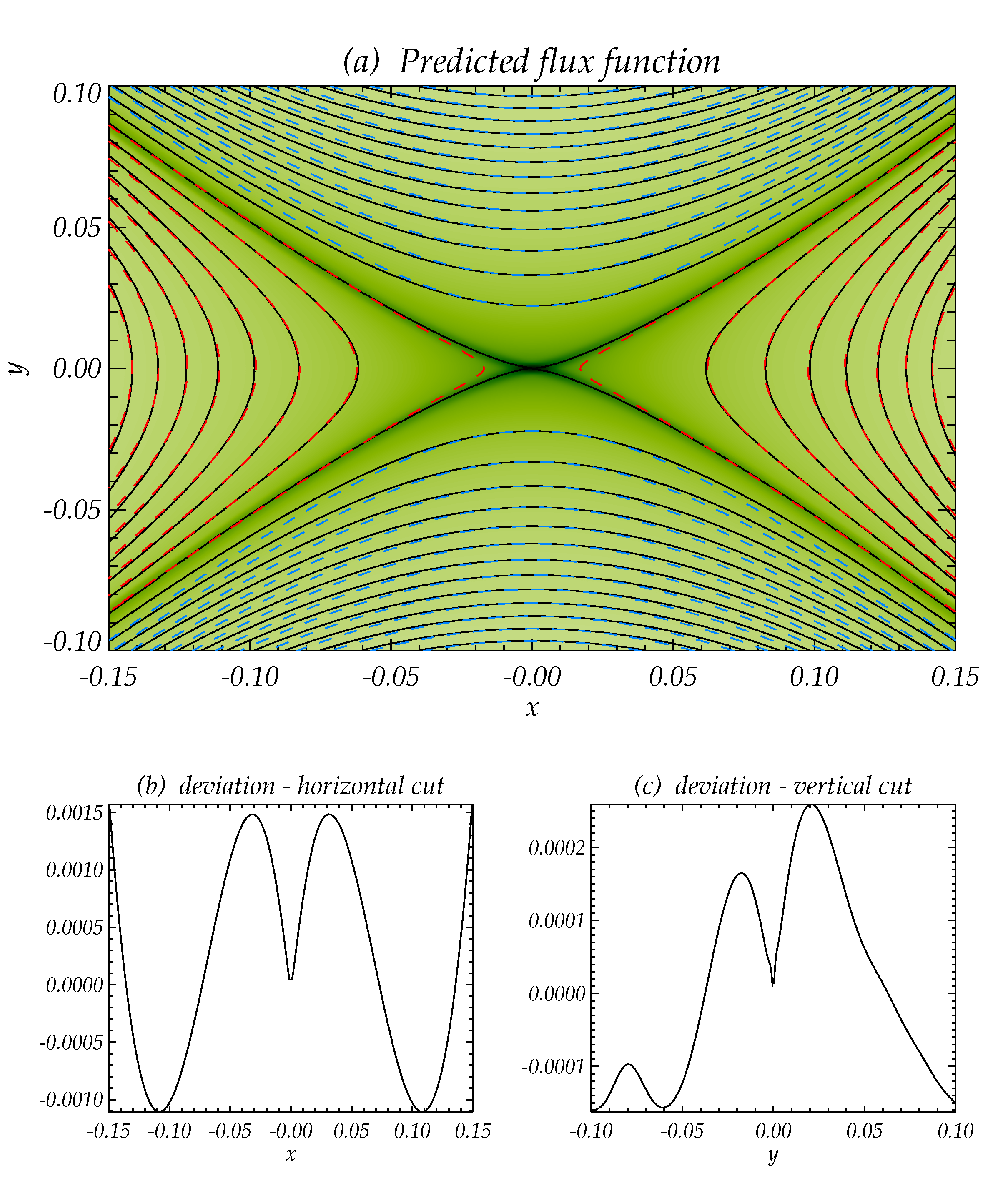}
\caption{Analytical prediction of the flux function, compared to numerical equilibrium state, for the experiment with $h=0.7$ and $p_0=0.375$. In (a) we show the numerical magnetic field lines (solid black) over a contour map of the current density, with the analytical contours of $A_z$ for inside the cusp (red dashed) and outside the cusp (blue dashed). In (b) and (c) we show the relative error of the analytical prediction from the numerical state for a horizontal cut (inside cusp) and a vertical cut (outside cusp) respectively.}
\label{fig:overplot}
\end{figure}

We suggest a poloidal flux function of the form
\begin{equation}
A_z(r,\theta)=A_0(r)+A_1(r,\theta) \;, \label{poloidal0}
\end{equation}
where
\begin{eqnarray}
A_0(r)&=&ar^p\;, \label{poloidal1} \\
A_1(r,\theta)&=&br^q\cos{2\theta}\;. \label{poloidal2}
\end{eqnarray}
This form can be compared to Eq. (\ref{fit}), using $\nabla^2 A_z=-j_z$. From Equations (\ref{poloidal0}), (\ref{poloidal1}) and (\ref{poloidal2}), we get
\begin{equation}
\nabla^2 A_z=ap^2r^{p-2} + b(q^2-4)r^{q-2}\cos{2\theta}
\end{equation}
Now, assuming the singular form for the flux function given by Eq. (\ref{sing}), combined with Eq. (\ref{poloidal0}), we obtain three relations between the parameters $a$, $p$, $q$ and the parameters $m$ and $n$, as follows,
\begin{eqnarray}
p&=&\frac{2}{1+n} \;, \\
-a&=&\left(\frac{j_{max}}{m^np^2}\right)^{1/(n+1)} \;, \\
q&=&\left(4+\frac{nj_{max}}{m^na^{n+1}}\right)^{1/2} \;.
\end{eqnarray}
Therefore, the only free parameters of our analytical fits are $(m,n,b)$. In Fig. \ref{fig:overplot}a, we show the contours of the numerical $A_z$, compared to the contours of our analytical form of $A_z(r,\theta)$. The fit is fairly good save at the origin and near the separatrices, where the assumptions above are not valid, since the field is not singular there. Figures \ref{fig:overplot}b and \ref{fig:overplot}c show the relative deviation of the analytical solution from the numerical in horizontal and vertical cuts at the center of the box respectively. The fits along these cuts are very good, with a very small error of the order of $10^{-3}-10^{-4}$. The parameters of the fit are summarised in Table \ref{tab:par}, for the experiment with $h=0.7$ and $p_0=0.375$, and with $j_{max}=8.5$. The regions inside and outside the cusp are denoted with $i$ and $o$ respectively.

These parameters depend on the initial plasma pressure and the initial current density, but also on the time we choose for our quasi-static equilibrium (i.e. they depend on the choice of the peak current). However, they provide a qualitative result for an non-force-free X-point equilibrium, shortly before magnetic reconnection occurs.

\begin{table}[ht]
\centering
\caption{Parameters of the analytical solutions for the regions inside ($i$) and outside ($o$) the cusp.}
\begin{tabular}{c|cc|c|ccc}
 & $m$ & $n$ & $b$ & $a$ & $p$ & $q$ \\
\hline
$i$ & $55\times10^4$ & 0.253 & 0.60 & $-0.18$ & 1.60 & 1.83 \\
$o$ & $17\times10^4$ & 0.294 & 0.66 & $-0.17$ & 1.55 & 1.82
\end{tabular}
\label{tab:par}
\end{table}


\section{Summary and conclusions}\label{sec:sec6}

We have presented evidence that the ideal MHD relaxation of two-dimensional magnetic X-points embedded in non-zero beta plasmas leads to a non-force-free quasi-static state that depends highly on the initial values of the background plasma pressure and electric current density. In this state, the system is in equilibrium everywhere save at the null and along the separatrices. Hence, the evolution of the system can be divided in two parts: (1) the quick viscous relaxation of the whole system, plus (2) an asymptotic regime with a very slow evolution of the current density at the null and along the four separatrices.

The slow increase of the current density at the location of the null is associated with the evolution to a singularity in an infinite amount of time. This slow evolution at the null makes the current accumulation region shorter and thinner as time elapses, concentrating the current at the exact location of the null, and also along the four separatrices. It should be noted that the large numerical resolution used here permits us to resolve the current layer both in width and length. However, note also that a singular value of the current will always be impossible to reach because (i) the singularity does not form in a finite time and (ii) any grid, no matter how fine, will eventually result in a numerical diffusion of the current.

Our overall results are completely different to those of \citet{Rastatter94}, \citet{Craig05} and \citet{Pontin05} for two main reasons: First, as time gets large, the system is not trying to collapse to a current sheet with a finite length, but it is concentrating all the current at the location of the null and along the separatrices. Second, this infinite current state is not achievable, and therefore, we must describe a final state in a quasi-static equilibrium at a given time, in which the current density has accumulated at the location of the null in a layer with finite length and width, that we are able to resolve in our numerical experiments. These results suggest that the idea of an infinitesimally thin current sheet must be then put aside when describing the MHD evolution of the system. The final state described in this paper may be identified with a moment shortly before magnetic reconnection takes place in a consistent non-force-free two-dimensional model.

We propose a functional form of the quasi-static equilibrium as $j_z(A_z)=j_{max}(1+mA_z)^{-n}$, and we derive a form for the flux function of $A_z(r,\theta)=ar^p+br^q\cos{2\theta}$, where the parameters $a$, $p$ and $q$ are well defined functions of $m$ and $n$. It is found that all these parameters are different from the regions inside and outside the cusp, and hence, the final equilibrium is different, and the system approaches the singularity in a different way for positive and negative values of $A_z$, i.e. inside and outside the cusp, respectively. Also, the final equilibrium depends strongly on the initial plasma pressure and current density. None of the above expressions are valid along the four separatrices, where $A_z=0$ but $j_z\ne j_{max}$. Hence, extra considerations may be needed for a full description in those regions.

It is found, both numerically and analytically, that an increase in the initial plasma pressure, and hence, an increase in the plasma beta, acts on the field by reducing both the strength of the singularity and its growth rate with time. This suggests \citep[see][]{Craig05} that plasma pressure acts against fast reconnection.

Even if this paper describes only a qualitative analysis of a 2D hydromagnetic equilibrium, it describes a fair approximation of the behaviour of the final state as the values of the initial plasma pressure and current density are varied. These two-dimensional contexts are of high relevance for systems with translational or rotational symmetries, and their study is useful for some astrophysical environments which can be well approximated by these properties of symmetry.

In a follow-up paper, we will evaluate current accumulations in three-dimensional equilibria which contain 3D magnetic null points. The characteristics of these environments are going to be completely different to the two-dimensional case, and the dynamical evolutions are less restrictive in the sense that the plasma has freedom to move in all three spatial directions.


\section*{Acknowledgments}

The authors would like to thank Profs D.W. Longcope and E.R. Priest for useful discussions. This work has been supported by the SOLAIRE European Training Network. Computations were carried out on the UKMHD consortium cluster funded by STFC and SRIF.


\bibliographystyle{aa}
\bibliography{aa17156.bib}


\end{document}

%% file: math_def.tex
\newcommand{\beq}{ \begin{eqnarray} }
\newcommand{\eeq}{ \end{eqnarray} }

\newcommand{\boldnabla}{\mbox{\boldmath$\nabla$}}

%% file: aa17156.bbl
\begin{thebibliography}{32}
\expandafter\ifx\csname natexlab\endcsname\relax\def\natexlab#1{#1}\fi

\bibitem[{{Antiochos} {et~al.}(1999){Antiochos}, {DeVore}, \&
  {Klimchuk}}]{Antiochos99}
{Antiochos}, S.~K., {DeVore}, C.~R., \& {Klimchuk}, J.~A. 1999, \apj, 510, 485

\bibitem[{{Arber} {et~al.}(2001){Arber}, {Longbottom}, {Gerrard}, \&
  {Milne}}]{Arber01}
{Arber}, T.~D., {Longbottom}, A.~W., {Gerrard}, C.~L., \& {Milne}, A.~M. 2001,
  J. Comp. Phys., 171, 151

\bibitem[{{Birn} {et~al.}(2003){Birn}, {Schindler}, \& {Hesse}}]{Birn03}
{Birn}, J., {Schindler}, K., \& {Hesse}, M. 2003, Journal of Geophysical
  Research (Space Physics), 108, 1337

\bibitem[{{Biskamp}(1986)}]{Biskamp86}
{Biskamp}, D. 1986, in Magnetic Reconnection and Turbulence, ed. {M.~A.~Dubois,
  D.~Gr{\'e}sellon, \& M.~N.~Bussac}, 19

\bibitem[{{Bungey} \& {Priest}(1995)}]{Bungey95}
{Bungey}, T.~N. \& {Priest}, E.~R. 1995, \aap, 293, 215

\bibitem[{{Craig}(1994)}]{Craig94}
{Craig}, I.~J.~D. 1994, \aap, 283, 331

\bibitem[{{Craig} \& {Henton}(1994)}]{Craig94a}
{Craig}, I.~J.~D. \& {Henton}, S.~M. 1994, \apj, 434, 192

\bibitem[{{Craig} \& {Litvinenko}(2005)}]{Craig05}
{Craig}, I.~J.~D. \& {Litvinenko}, Y.~E. 2005, Phys. Plasmas, 12, 032301

\bibitem[{{Dungey}(1953)}]{Dungey53}
{Dungey}, J.~W. 1953, \mnras, 113, 679

\bibitem[{{Forbes} {et~al.}(2006){Forbes}, {Linker}, {Chen}, {Cid}, {K{\'o}ta},
  {Lee}, {Mann}, {Miki{\'c}}, {Potgieter}, {Schmidt}, {Siscoe}, {Vainio},
  {Antiochos}, \& {Riley}}]{Forbes06}
{Forbes}, T.~G., {Linker}, J.~A., {Chen}, J., {et~al.} 2006, \ssr, 123, 251

\bibitem[{{Friedel} {et~al.}(1996){Friedel}, {Grauer}, \&
  {Marliani}}]{Friedel96}
{Friedel}, H., {Grauer}, R., \& {Marliani}, C. 1996, Computer, 8001

\bibitem[{{Fuentes-Fern{\'a}ndez} {et~al.}(2010){Fuentes-Fern{\'a}ndez},
  {Parnell}, \& {Hood}}]{Fuentes10}
{Fuentes-Fern{\'a}ndez}, J., {Parnell}, C.~E., \& {Hood}, A.~W. 2010, \aap,
  514, A90

\bibitem[{{Furth} {et~al.}(1973){Furth}, {Rutherford}, \& {Selberg}}]{Furth73}
{Furth}, H.~P., {Rutherford}, P.~H., \& {Selberg}, H. 1973, Physics of Fluids,
  16, 1054

\bibitem[{{Grauer} \& {Marliani}(1998)}]{Grauer98}
{Grauer}, R. \& {Marliani}, C. 1998, Physics of Plasmas, 5, 2544

\bibitem[{{Green}(1965)}]{Green65}
{Green}, R.~M. 1965, in Stellar and Solar Magnetic Fields, ed. R.LustI, AU
  Symp. 22 p.389

\bibitem[{{Hesse} \& {Schindler}(2001)}]{Hesse01}
{Hesse}, M. \& {Schindler}, K. 2001, Earth, Planets, and Space, 53, 645

\bibitem[{{Klapper}(1997)}]{Klapper97}
{Klapper}, I. 1997, Physics of Plasmas, 5, 910

\bibitem[{{Longcope} \& {Parnell}(2009)}]{Longcope09}
{Longcope}, D.~W. \& {Parnell}, C.~E. 2009, \solphys, 254, 51

\bibitem[{{McLaughlin} \& {Hood}(2004)}]{McLaughlin04}
{McLaughlin}, J.~A. \& {Hood}, A.~W. 2004, \aap, 420, 1129

\bibitem[{{McLaughlin} \& {Hood}(2006)}]{McLaughlin06}
{McLaughlin}, J.~A. \& {Hood}, A.~W. 2006, \aap, 459, 641

\bibitem[{{Parker}(1957)}]{Parker57}
{Parker}, E.~N. 1957, \jgr, 62, 509

\bibitem[{{Petschek}(1964)}]{Petschek64}
{Petschek}, H.~E. 1964, NASA Special Publication, 50, 425

\bibitem[{{Pontin} \& {Craig}(2005)}]{Pontin05}
{Pontin}, D.~I. \& {Craig}, I.~J.~D. 2005, Phys. Plasmas, 12, 072112

\bibitem[{{Priest} \& {Forbes}(1986)}]{Priest86}
{Priest}, E.~R. \& {Forbes}, T.~G. 1986, \jgr, 91, 5579

\bibitem[{{Rast\"atter} {et~al.}(1994){Rast\"atter}, {Voge}, \&
  {Schindler}}]{Rastatter94}
{Rast\"atter}, L., {Voge}, A., \& {Schindler}, K. 1994, Phys. Plasmas, 1, 3414

\bibitem[{{Shibata} {et~al.}(1993){Shibata}, {Nozawa}, \&
  {Matsumoto}}]{Shibata93}
{Shibata}, K., {Nozawa}, S., \& {Matsumoto}, R. 1993, in Astronomical Society
  of the Pacific Conference Series, Vol.~46, IAU Colloq. 141: The Magnetic and
  Velocity Fields of Solar Active Regions, ed. {H.~Zirin, G.~Ai, \& H.~Wang},
  500

\bibitem[{{Somov} \& {Syrovatskii}(1976)}]{Somov76}
{Somov}, B.~V. \& {Syrovatskii}, S.~I. 1976, Neutral Current Sheets on Plasmas,
  13

\bibitem[{{Sweet}(1958)}]{Sweet58}
{Sweet}, P.~A. 1958, in IAU Symposium, Vol.~6, Electromagnetic Phenomena in
  Cosmical Physics, ed. {B.~Lehnert}, 123

\bibitem[{{Vainshtein}(1990)}]{Vainshtein90}
{Vainshtein}, S.~I. 1990, \aap, 230, 238

\bibitem[{{Vekstein} \& {Priest}(1993)}]{Vekstein93}
{Vekstein}, G. \& {Priest}, E.~R. 1993, \solphys, 146, 119

\bibitem[{{Zuccarello} {et~al.}(2009){Zuccarello}, {Jacobs}, {Soenen},
  {Poedts}, {van der Holst}, \& {Zuccarello}}]{Zucarello09}
{Zuccarello}, F.~P., {Jacobs}, C., {Soenen}, A., {et~al.} 2009, \aap, 507, 441

\bibitem[{{Zwingmann} {et~al.}(1985){Zwingmann}, {Schindler}, \&
  {Birn}}]{Zwingmann85}
{Zwingmann}, W., {Schindler}, K., \& {Birn}, J. 1985, Solar Physics, 99, 133

\end{thebibliography}
